\documentclass{emulateapj-rtx4}
\shorttitle{NSV 11749, an elder sibling of V605 Aql and V4334 Sgr?}
\shortauthors{M. M. Miller Bertolami, R. D. Rohrmann, A. Granada \& L. G.  Althaus}
\begin{document}
\title{NSV 11749, an elder sibling of the born again stars V605 Aql and V4334 Sgr?}
\author{M. M. Miller Bertolami$^{1,2}$, R. D. Rohrmann$^{3}$, A. Granada$^4$  \& L. G.  Althaus$^{1,2}$}
\affil{$^{1}$Facultad de Ciencias Astron\'omicas y Geof\'isicas, Universidad Nacional de La Plata, Paseo del Bosque s/n, 1900 La Plata, Argentina.\\
$^{2}$CCT-La Plata, CONICET, Argentina.\\
$^{3}$Instituto de Ciencias Astron\'omicas, de la Tierra y del Espacio, CONICET, Av. de Espa\~na 1512 (Sur) CC 49,5400 San Juan, Argentina.\\
$^{4}$Observatoire Astronomique de l'Universit\'e de Gen\`eve 51, Chemin des Maillettes, CH-1290, Sauverny, Suisse.}
\email{mmiller@fcaglp.unlp.edu.ar}

\begin{abstract}

We argue that NSV 11749, an eruption observed in the early twentieth
century, was a rare event known as ``very late thermal pulse'' (VLTP).
To support our argument we compare the lightcurve of NSV 11749 with
those of the two bonafide VLTP objects known to date, V4334 Sgr and
V605 Aql, and with those predicted by state of the art stellar
evolution models.  Next, we explore the IPHAS and 2MASS
catalogues for possible counterparts of the eruption.  Our analysis
shows that the VLTP scenario outperforms all other proposed scenarios
as an explanation of NSV 11749.  We identify an IPHAS/2MASS source
  at the eruption location of NSV 11749. The derived colors suggest
  that the object is not enshrouded in a thick dust shell as V605 Aql
  and V4334 Sgr. Also the absence of an apparent planetary nebula (PN)
  at the eruption location suggests differences with known VLTP
  objects which might be linked to the intensity of the eruption and
  the mass of the object. Further exploration of this source and
scenario seems desirable.
If NSV 11749 was a born again star, it would be the third event of its
kind to have been observed and will strongly help us to increase our
understanding on the later stages of stellar evolution and violent
reactive convective burning.

\end{abstract}


\section{Introduction}

About a fifth of the stars departing from the Asymptotic Giant Branch
(AGB) are expected to undergo a final thermal pulse during their
post-AGB evolution \citep{1984ApJ...277..333I}.  When this happens,
the pre-white dwarf is predicted to be temporarily reborn as a yellow
giant \citep{1979A&A....79..108S} in the so called ``Born Again AGB''
scenario \citep{1984ApJ...277..333I}. The transition from the
pre-white dwarf to the giant configuration is expected to be very
rapid; being of a few years in the very late thermal pulse (VLTP)
flavor \citep{1995LNP...443...48I} and of the order of a century in
the late thermal pulse (LTP) case \citep{1979A&A....79..108S}.  Due to
the short duration of these events, although 10\% of post-AGB stars
are expected to undergo a VLTP, they are extremely rare from an
observational perspective. Indeed, only two objects (V605 Aql and
V4334 Sgr) have been identified as stars undergoing a VLTP (see
\citealt{2000AJ....119.2360D, 2002Ap&SS.279..183D}), while a
  third has been identified with the LTP flavor of the scenario (FG
  Sge, see \citealt{2006A&A...459..885J} and references therein).
Although observationally rare, individual VLTP stars are extremely
valuable as they are key to understand the formation of C-rich
H-deficient stars, such as [WC]-CSPNe, PG1159 and RCrB (see
\citealt{2006PASP..118..183W} and \citealt{1996PASP..108..225C} for a
review) and the formation of H-deficient white dwarfs ---which
comprise about $\sim 20$\% of known white dwarfs. In addition, born
again star  events are a key test for our understanding of s-process
during the thermal pulse phase of the AGB
(\citealt{1999A&A...343..507A} and \citealt{2006A&A...459..885J}) and
also for understanding of reactive convective burning in the
interior of stars \citep{2011ApJ...727...89H}.

A rough estimate suggests that the birth rate of planetary nebulae in
our galaxy is of about $\sim 1$ every year
\citep{2002Ap&SS.279..171Z}. If 10\% of their central stars become
VLTP giants, then we should expect such events in our galaxy to take
place at a rate of about one per decade. Due to their high intrinsic
brightnesses ($M_v\sim -2 ... -4 $) these objects can be easily
detected at large distances within our galaxy. Hence, it should not be
strange if some born again eruptions are waiting to be identified in
old and new star surveys.  This could be the case for the object NSV
11749. After an excellent systematic study of all useful Harvard
plates, \cite{2005JAVSO..34...43W} was able to reconstruct the
outburst lightcurve of this star. The plates show that the object was
fainter than $m_{\rm pg}\sim 15.5$ on 1897.6, became visible for the
first time at $m_{\rm pg}\sim 14$ on 1899.5 and reached maximum at
$m_{\rm pg}\sim 12.5$ on 1903.4. Then it remained at about maximum
brightness until 1907.6 when it started to decline, becoming
undetectable after 1911.6 ---with four possible detections at $m_{\rm
  pg}\sim 17$ some decades later. Finally, during the declining phase,
the star showed three sudden disappearances (fainter than $m_{\rm
  pg}\sim 14$) before finally fading into oblivion. In this letter we
argue that NSV 11749 was a VLTP event and explore for possible
counterparts in IPHAS and 2MASS catalogs.

\section{The lightcurve of NSV 11749}
\begin{figure}
\begin{center}
  \includegraphics[clip, angle=0, width=8cm]{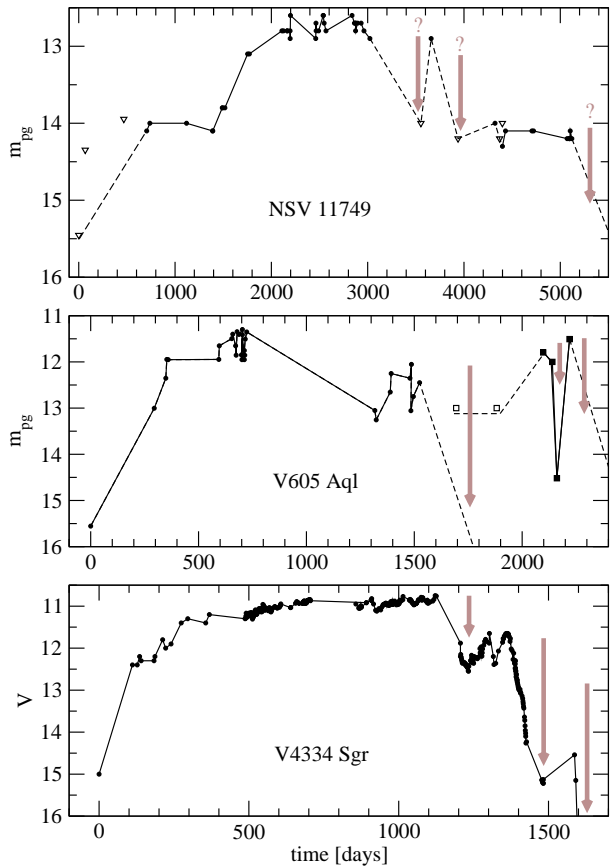}
\caption{Lightcurves of NSV 11749, V605 Aql and V4334 Sgr. Thick
  downwards arrows indicate RCrB-like sudden drops in brightness.}
\label{fig:curvas_3_objetos} 
\end{center}
\end{figure}
The lightcurve of NSV 11749 bears a strong resemblance with those of
VLTP-objects (V4334 Sgr and V605 Aql). As can be seen in
Fig. \ref{fig:curvas_3_objetos} NSV 11749 increased its brightness by
more than 2.5 magnitudes in the first 2000 days of its eruption
(i.e. $dm_{\rm pg}/dt> 0.45^m$/yr), stayed at a maximum brightness of
$m_{\rm pg}\sim 12.6^m$ for about 1000 days and then started to
experience sudden dimmings (of more than $1^m$) after finally
disappearing from view $\sim 13$ yr after its eruption. These three
sudden disappearances of the star are particularly worth noting as
they are very similar to those observed in the born again stars
(\citealt{2000AJ....119.2360D, 2002Ap&SS.279..183D}). At least in the
best studied born again star (V4334 Sgr,
\citealt{2000AJ....119.2360D}) it is clear that these sudden dimmings
are caused by carbon-dust ejection episodes similar to those observed
in R Coronae Borealis Stars (RCrB stars).  In
Fig. \ref{fig:curvas_3_objetos} we compare the lightcurve of NSV 11749
with those of the two known fast born again stars (V4334 Sgr and V605
Aql). The three lightcurves share the same main features: Namely, in a
period of years they show an outburst stage, a steady stage at maximum
brightness, a phase of RCrB-like declines and finally a complete
disappearance from view.  These similarities alone are a strong
argument in favor of a similar explanation for all three stars, and
thus for a born again (VLTP) explanation of the lightcurve of NSV
11749.  However, quantitatively, there is a significant difference in
the timescales of NSV 11749 and both V4334 Sgr or V605 Aql. In both
V4334 Sgr and V605 Aql the sequence of events happened at a faster
pace, increasing more than 3.5$^m$ during the first year. Also V4334
Sgr (V605 Aql) disappeared from view much faster, only $\sim 4.4$ yr
($\sim 6.3$ yr) after the eruption.

Then, the main question is to know if some born again stars could
evolve a few times slower than observed in V4334 Sgr or V605
Aql. Stellar evolution models (\citealt{2001ApJ...554L..71H},
\citealt{2007MNRAS.380..763M}) suggest that, indeed, that is the case.

\section{Theoretical VLTP lightcurves}
\begin{figure}
\begin{center}
  \includegraphics[clip, angle=0 , width=8.5cm]{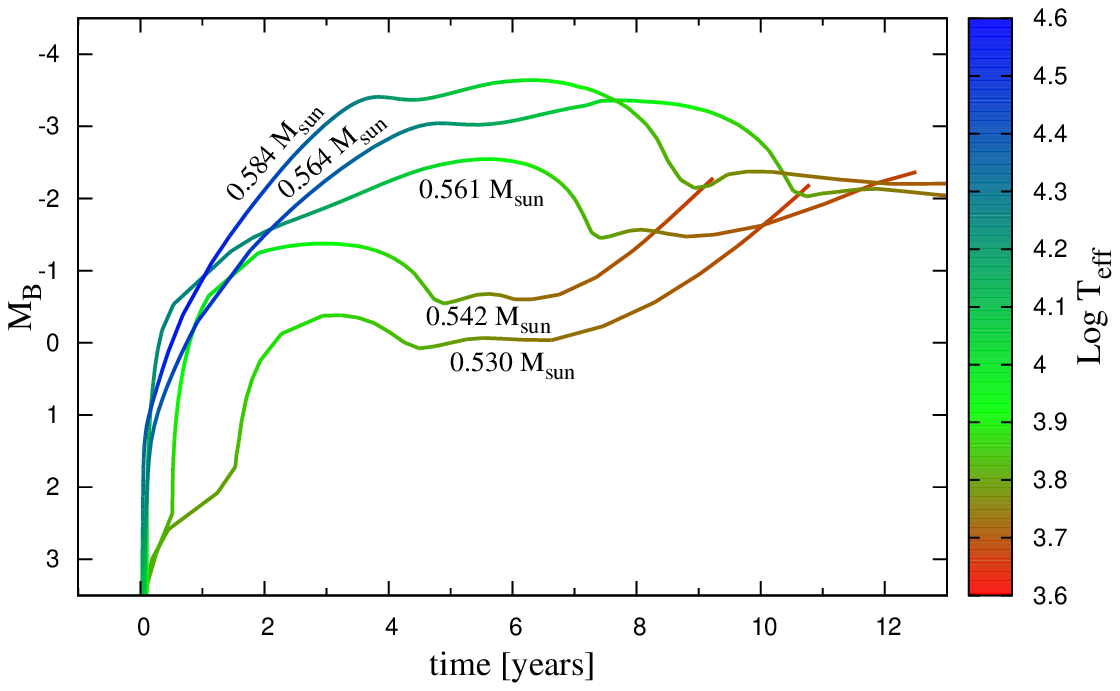}\\
\caption{Theoretical $M_B$ lightcurves for the fast-VLTP sequences of
  \cite{2007MNRAS.380..763M}.}
\label{fig:curvas_V_B} 
\end{center}
\end{figure}
According to stellar evolution models VLTP eruption lightcurves and
temperatures are dependent on the mass of the erupting star. Then, a
direct comparison of NSV 11749 lightcurve with those of the two known
VLTP objects (V605 Aql and V4334 Sgr) might be misleading.

In order to compare theoretical eruptions in VLTP models with the
observed lightcurve of NSV 11749 it is necessary to construct
theoretical $B$ and $V$ lightcurves. In the absence of bolometric
corrections for H-deficient stars in the wide range of temperatures
covered by VLTP eruptions, we have relied on theoretical model
atmospheres to predict the expected $B$ and $V$ magnitudes of the
stellar evolution models. Stellar atmospheres have been computed
within the assumption of plane parallel geometry and LTE, including
the opacities of all relevant major atoms (although without molecules
or dust). Although plane parallel geometry is not justified at the
very low surface gravities attained by the models some years after the
eruption, it is reasonable during the first years of the eruption,
i.e. before the development of the RCrB-stage in real
stars. Abundances in stellar atmosphere computations have been chosen
to reflect those predicted by VLTP models
(\citealt{2007MNRAS.380..763M}), but tests show that lightcurves would be very
similar if we had choosen abundances like those observed in
V4334 Sgr \citep{1999A&A...343..507A}.  Lightcurves computed with
these model atmospheres and the stellar evolution sequences of
\cite{2007MNRAS.380..763M} are shown in Fig.
\ref{fig:curvas_V_B}. Also shown is the effective temperatures
predicted by the models during the eruption.

As shown in Fig. \ref{fig:curvas_V_B}, our models predict different
lightcurves depending on the mass of the remnant that undergoes the
eruption. In particular note that after the fast initial rise, the
lightcurves either stall or slightly diminish.  After reaching a
plateau for a few years, the theoretical lightcurves start to rise
again as the sequences increase their luminosities without a strong
change in temperature. However, this happens after the temperature
falls below $\log T_{\rm eff}\sim 3.7$ and then we expect RCrB-like
events to develop.  In fact, the three identified born again stars (FG
Sgr, V605 Aql and V4334 Sgr) have shown RCrB extinction episodes after
the temperature fell below $\log T_{\rm eff}\sim 3.7$
(\citealt{2006A&A...459..885J}, \citealt{1997AJ....114.2679C} and
\citealt{2000AJ....119.2360D}).  Thus, synthetic lightcurves will not
reflect the observed behavior from this point onwards.  Then, the
intrinsic maximum brightness of fast-VLTP sequences, after the fast
rise in brightness, spans a wide range from $\sim -1$ to $-4$ both in
V and B bands.

\subsection{Test: Comparison with V605 Aql and V4334 Sgr}
\begin{figure}
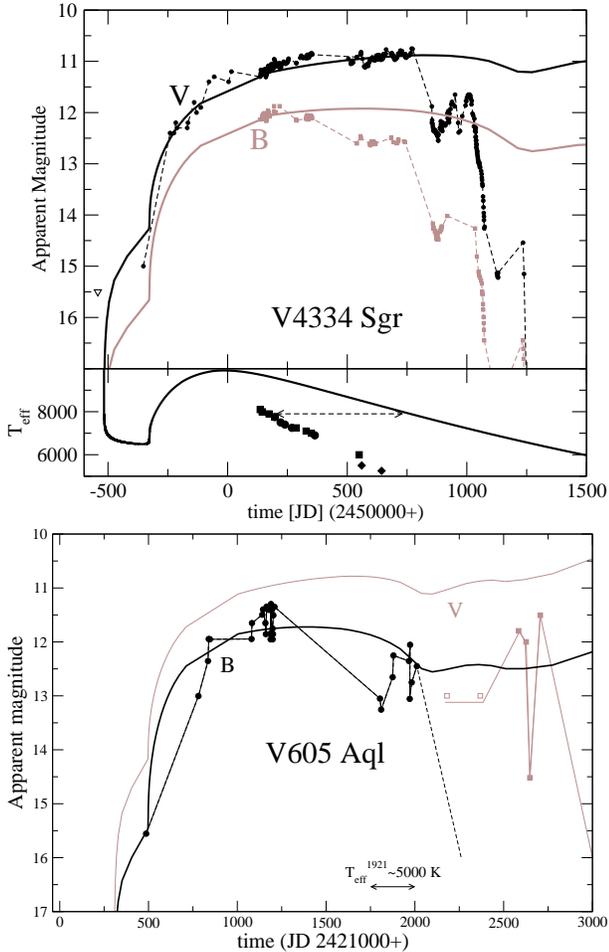

\begin{center}
 \includegraphics[clip, , width=8cm]{fig3u.eps}\\
 \includegraphics[clip, , width=8cm]{fig3l.eps}
\caption{V and B lightcurves of V4334 Sgr (upper panel) and V605 Aql
  (lower panel) compared with our most similar lightcurve (sequence
  $0.542$$M_{\odot}$\ of \citealt{2007MNRAS.380..763M}).}
\label{fig:V4334Sgr_V605Aql} 
\end{center}
\end{figure}
In order to understand to which extent our lightcurves can be trusted
when comparing with real stars, we now compare our lightcurves with
those of the two bonafide VLTP objects, V605 Aql and V4334 Sgr. In
Fig. \ref{fig:V4334Sgr_V605Aql} we compare the lightcurve of our best fit
model lightcurve (0.542$M_{\odot}$) with the visual lightcurve of V4334 Sgr
(\citealt{1997Takami}, \citealt{1997AJ....114.1657D,
  2000AJ....119.2360D}) and  with the
photographic lightcurve of V605 Aql published by
\cite{2002Ap&SS.279..183D}. To compare $B$ and $m_{\rm pg}$
magnitudes, we adopt $m_{\rm pg}=B+0.11$ as in
\cite{1997AJ....114.2679C}. 
The similarities between the predicted and observed lightcurves are
apparent. It is worth noting that no tuning of the theoretical models
has been carried out in order to fit the observed
lightcurves. Fig. \ref{fig:V4334Sgr_V605Aql} just
displays our sequence with the most similar aspect with the observed
ones.

Assuming the interstellar extinction model of
\cite{1997AJ....114.2043H}, we find for V4334 Sgr ($V-M_V\sim 12.7^m$,
for our most similar lightcurve) a distance of $d\sim 1.6$ kpc and
$A_V\sim 1.9^m$. Interestingly enough, this value is within the
recommended values by \cite{2002Ap&SS.279...79K},
$d=2^{+1}_{-0.6}$ kpc, on the basis of several independent distance
determinations. Thus our 0.542$M_{\odot}$\ lightcurve not only predicts a
correct lightcurve shape for V4334 Sgr but also a correct absolute
magnitude. Also, at the distance of $d\sim 1.6$ kpc, and a derived
value of $A_B\sim 2.5^m$ our model also predicts the maximum
brightness in the $B$ band (see Fig. \ref{fig:V4334Sgr_V605Aql}).
On the other hand, it is clear from Fig. \ref{fig:V4334Sgr_V605Aql} that our
model is not able to predict simultaneously the correct luminosity and
temperature evolution (although its cooling speed, $dT_{\rm eff}/dt$,
is very similar to that of V4334 Sgr). Also the
0.542$M_{\odot}$\ pre-outburst location in the HR diagram might be at
variance with a possible pre-discovery detection of V4334 Sgr's
progenitor in 1976 (see \citealt{2007MNRAS.380..763M}).

For the case of V605 Aql ($B-M_B\sim 13.1^m$, for our most similar
lightcurve) we obtain a distance of $d\sim 1.9$ kpc (and $A_B\sim 2^m$,
assuming $R=A_V/(A_B-A_V)=3.1$), a value significantly lower than
derived in previous works; 2.7 kpc$<d<$6.0 kpc
\citep{1997AJ....114.2679C}.  However, it has to be kept in mind that
larger distances would have been obtained if the extinction is not as
high as suggested by \cite{1997AJ....114.2043H} in that particular
direction of the sky ($30^\circ<\lambda<40^\circ$).

From these comparisons we conclude that our theoretical lightcurve
shapes are very similar to those observed in real born again stars and
can be used to identify born again star candidates. We notice, however, that
theoretical models are not able to fit, simultaneously, all observed
features. This is most probably due to the uncertainties in the
treatment of the violent reactive convective burning of H in the
models, something that it is still badly understood
\citep{2011ApJ...727...89H}.

\subsection{Comparison with NSV 11749}
\begin{figure}
\begin{center}
  \includegraphics[clip, , width=8cm]{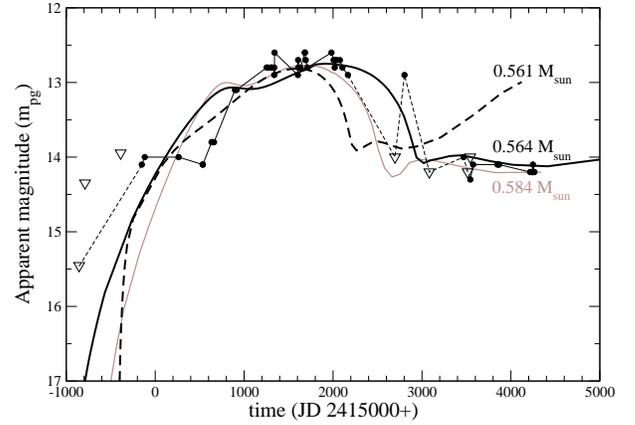}
\caption{Lightcurve of NSV 11749 compared with some theoretical B
  lightcurves.}
\label{fig:NSV11749} 
\end{center}
\end{figure}
In Fig. \ref{fig:NSV11749} we compare B-lightcurves of our 0.561,
0.564 and 0.584$M_{\odot}$\ sequences with the photographic lightcurve
reconstructed by \cite{2005JAVSO..34...43W}. As can be seen, the
theoretical lightcurves account for the brightening speed observed in
NSV 11749 of about $\sim 0.5$ magnitudes per year. Also, our
lightcurves predict that the star will stay at maximum brightness for
a few years. More interesting NSV 11749 has shown three sudden
extinctions between 1907 and 1910 before disappearing from Harvard
plates in 1912.  These extinctions occur when our sequence shows a
temperature $\log T_{\rm eff }\lesssim 3.8$ and thus when real born
again stars have shown us that RCrB-like extinction events are
expected to occur.  Thus, our sequences not only reproduce the
eruption lightcurve but also agree with the interpretation of the
three drops in brightness observed in NSV 11749 as being caused
by RCrB-like events.  Then, our models show that while NSV 11749 has
increased its brightness by a factor about 2 to 4 slower than V4334 Sgr
or V605 Aql, its lightcurve is well within the expected behavior for
VLTP eruptions of different masses.  Comparing the absolute magnitudes
of the theoretical models with those observed in NSV 11749, $B-M_B\sim
16^m$ for our 0.564$M_{\odot}$\ sequence, we roughly estimate a distance of
$d\sim 3.2$ kpc and $A_B\sim 4.2$. Had we compared with our
0.561$M_{\odot}$\ (0.584$M_{\odot}$) sequence, for which $B-M_B\sim 15.25^m$
($B-M_B\sim 16.3^m$), we would have estimated a distance of $\sim 2.3$
 kpc ($\sim 3.6$ kpc) and $A_B\sim 4$ ($A_B\sim 4.3$).  It must be
noted that these distance estimates are very uncertain as they not
only depend on the accuracy of born again models but also
distances could be much larger if $A_B$ is overestimated by
\cite{1997AJ....114.2043H}.

\section{Possible present counterpart of NSV 11749}
%
%
\begin{figure}
\begin{center}
  \includegraphics[clip, , width=8.3cm]{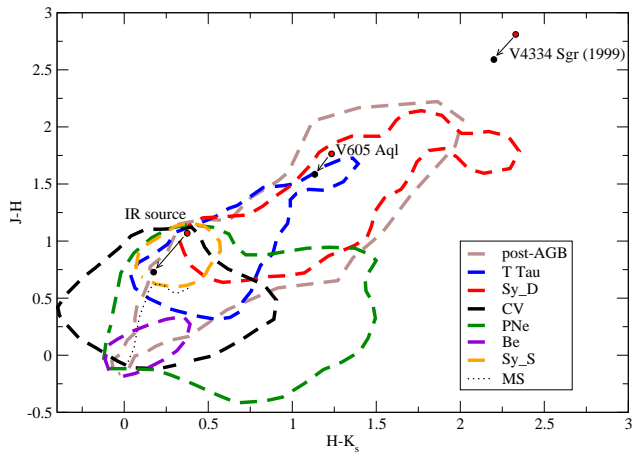}
\caption{2MASS color-color diagram for the source at coordinates of
  NSV 11749 compared with the 2MASS colors of known VLTP objects. Also
  shown are sketchs of the location expected for different types of
  objects according to \cite{2009A&A...504..291V, 2009A&A...502..113V}. Red dots indicate observed colors and black dots dereddened colors.}
\label{fig:colores} 
\end{center}
\end{figure}

Prompted by the strong resemblance of NSV 11749 with born again
lightcurves (real and theoretical) we looked into 2MASS
\citep{2003tmc..book.....C} and IPHAS \citep{2008ASPC..394..197G}
catalogs for possible counterparts. If NSV 11749 experienced a born
again event $\sim$100 yr ago we would expect it to be, by now, either
enshrouded in a thick dust shell similar to post-AGB stars or
reheating as a new central star of a planetary nebulae (as seen in
V605 Aql).  In order to constrain our search for present counterparts
coordinates for NSV 11749 have been redetermined by the Digital Access
to a Sky Century at Harvard (DASCH, \citealt{2009ASPC..410..101G})
team from 6 plates from the Harvard College University plate archive.
These plates were scanned and analized with the DASCH photometry
pipeline, which yielded coordinates $\alpha=19^h07^m42.41^s$ and
$\delta=00^\circ02'51.4''$ with a RMS error of $\sigma\sim
1''$\footnote{These coordinates are remarkably close to those
  suggested by \cite{2005JAVSO..34...43W} which are several arcmins
  away from those recorded in the NSV catalog. This is because NSV
  coordinates correspond to Luyten's published discovery position,
  that was only estimated from grids traced over the plate (Williams
  private communication).}. Their photometry was also consistent with
that presented by \cite{2005JAVSO..34...43W}.

Only one infrared source (from now on IRS) very near to the location
of NSV 11749, $\alpha=19^h07^m42.4^s$ and $\delta=00^\circ02'51.0''$,
is within $3\sigma$ from the derived coordinates. The IRS is included
in both 2MASS and IPHAS catalogs with magnitudes $J=10.794$,
$H=9.726$, $K_s=9.351$ (2MASS) and $r'=14.509$, $i'=13.053$,
$H_\alpha=12.994$ (IPHAS). As shown in Fig. \ref{fig:colores}, 2MASS
colors for the IRS are consistent with those of symbiotic stars,
T-Tauri stars, cataclismic variables, post-AGB stars and PNe. Also,
dereddened colors (\citealt{1985ApJ...288..618R}, assuming $d=3.2$
kpc) fall very close to main sequence stars. Fortunately, IPHAS colors
for NSV 11749 fall above the cut defined by \cite{2009A&A...504..291V}
to isolate emision line objects (Zone 2, see Fig. 1 of
\citealt{2009A&A...504..291V}) and we can discard a main sequence
star. Also, IPHAS colors fall in a region of the color-color diagram
populated by symbiotic stars, T-Tauri stars and PNe but away from
post-AGB stars (the IRS has higher $H_\alpha$ brightness). Note,
however, that despite the similar IPHAS and 2MASS colors, a FU Ori
(i.e. T-Tauri) or symbiotic nova (i.e. symbiotic star) explanation for
NSV 11749 is unlikely (see next section). In Fig. \ref{fig:colores} we
also compare the 2MASS colors of the IRS with those derived for V4334
Sgr and V605 Aql ---dereddened assuming recomended distances of $d=2$
kpc and $d=3.5$ kpc respectively. As it is apparent, 2MASS colors for
the three objects are very different. The IRS is bluer than dust
enshrouded symbiotic stars (Sy\_D) and dust enshrouded VLTP
objects. Thus, 2MASS colors suggest that the IRS is not strongly
enshrouded by dust. Also the IRS is much brighter than the present
state of V605 Aql, which is suspected to be completely hidden behind a
thick dust torus \citep{2006ApJ...646L..69C} but similar to the
brightness of V4334 Sgr before the beginning of dust extinction events
($J\sim 9.5 ...7$, \citealt{2000AstL...26..506T}).

Finally, an old PN would be expectable within the born again scenario,
as all previous born again objects (FG Sge, V605 Aql and V4334 Sgr)
show such PNe. Inspection of UKST and IPHAS images around NSV 11749 do
not reveal any PN around the eruption.  However, as the formation of a
PN depends on the evolutionary speed of its central star, the absence
of a PN could be just the consequence of a different mass of the
erupting star (as already suggested by its lightcurve). Finally, note
that, material ejected during the eruption (assuming 100 km/s as in
V605 Aql, \citealt{1997AJ....114.2679C}) would be smaller than 1" and,
thus, not resolved. If NSV 11749 was a VLTP, the progenitor star had to
be different from those of V605 Aql and V4334 Sgr as the object does
not seem to be now surrounded by a PN or enshrouded in a thick dust
shell.

\section{Discussion and final remarks.}
Based on its photometric lightcurve, Williams suggested two possible
scenarios to explain NSV 11749, either a slow nova or a FU Ori type
star.  In particular a FU Ori event would be consistent with the 2MASS
colors of the IRS that show it similar to T-Tauri stars. However as
already mentioned by \cite{2005JAVSO..34...43W} these scenarios are
unable to account both the brigthness increase and dimming. While slow
novae decline in timescales of the order of a year
\citep{2003MNRAS.344.1219H} in qualitative agreement with NSV 11749,
their rising is much faster increasing more than 5 magnitudes in a few
days. The opposite happens with the FU Ori scenario. While typical FU
Ori stars increase their brightness in a period of the order of a year
their dimming is extremely slow, declining only a few magnitudes in
decades \citep{1996ARA&A..34..207H}. Then both scenarios fail to mach
the observed behavior of NSV 11749 with both rising and dimming
taking place in timescales of years.

A third alternative scenario suggested by the slow rising lightcurve
is that of a symbiotic nova (also known as very slow novas, see
\citealt{2010arXiv1011.5657M} for a review). Symbiotic novae are
thermonuclear novae that take place in symbiotic binary systems that
would allow a natural link with the IRS identified in the previous
section. The eruption period of these objects can last from month to
years, thus naturally accounting for NSV 11749 observed eruption
lightcurve. However the decline of these eruptions is extremely slow
lasting for decades and even centuries (e.g. PU Vul). Thus, as in the
case of the two previous scenarios, no simultaneous agreement with the
eruption and declining timescales can be achieved. 

Clearly, the born again scenario outperforms all other proposed
explanations for NSV 11749.  In fact, as discussed in Sect. 2, the
qualitative agreement between the lightcurves of NSV 11749 and known
VLTP objects is very good.  The most significant quantitative
difference is that NSV 11749 eruption was a few times ($\sim 5$ times)
slower with the rising taking place in about 5 years. However, we have
shown in previous sections that these differences are expectable from
differences in the mass of the star and, even better, that NSV 11749
lightcurve can be quantitatively reproduced by VLTP models. In fact,
differences in mass of the star could be related with the absence of a
PN around the eruption.

With the aid of synthetic born again lightcurves we have presented
strong arguments in favour of a VLTP explanation for NSV 11749. If
this is so, NSV 11749 has some differences with both V4334 Sgr and
V605 Aql (no PN, probably not enshrouded by dust). This finding will
strongly increase our understanding of the late stages of stellar
evolution. In particular it will boost our understanding of the
formation of H-deficient stars and of the reactive-convective burning
phases of stellar evolution.

\acknowledgments 

This work has been supported by grants PIP 112-200801-00904 and
PICT-2010-0861 from CONICET and ANCyT, respectively. The authors thank
D. B. Williams for very helpful comments and data and also an
anonymous referee for his suggestion to obtain new coordinates for NSV
11749 and contact J. Grindlay which strongly improved the final
version of the manuscript. J. Grindlay, A. Doane and E. Los and the
DASCH project are gratefully acknowledged for scanning 10 Harvard
A-plates around the time of the outburst and deriving magnitudes
and the stellar position presented here. DASCH is supported by NSF grant
AST-0909073 and the {\it Cornel and Cynthia K.  Sarosdy Fund for
  DASCH}. This research has made an extensive use of NASA's
Astrophysics Data System.


\end{document}